\documentclass[11pt,a4paper]{article}

\usepackage[a4paper,text={150mm,240mm},centering,headsep=10mm,footskip=15mm]{geometry}
\usepackage[T1]{fontenc}
\usepackage{amsmath,amssymb}
\usepackage{amsfonts}
\usepackage{url}
\usepackage{mathrsfs} 

\newcommand{\D}{\dagger}

\title{On the description of subsystems in relativistic hypersurface Bohmian mechanics}

\author{
Detlef Dürr\thanks{duerr@math.lmu.de} \ and
Matthias Lienert\thanks{Corresponding author, lienert@math.lmu.de}\\[0.2cm]
	Mathematisches Institut, Ludwig-Maximilians-Universit\"at\\
	Theresienstr. 39, 80333 M\"unchen, Germany
}

\date{September 10, 2014}

\begin{document}

\maketitle

\begin{abstract}
	\noindent
	A candidate for a realistic relativistic quantum theory is the hypersurface Bohm-Dirac model. Its formulation uses a foliation of spacetime into space-like hypersurfaces. In order to apply the theory and to make contact with the usual quantum formalism one needs a framework for the description of subsystems. The presence of spin together with the foliation renders the subsystem description more complicated as compared to the non-relativistic case with spin. In this paper, we provide such a framework in terms of an appropriate conditional density matrix and an effective wave function as well as clarify their relation, thereby generalizing previous subsystem descriptions in the non-relativistic case.\\

\end{abstract}

\section{Introduction}
The possibility of describing a subsystem of a larger system in an autonomous way is basic to physics. In quantum physics entanglement prevails and an autonomous subsystem description seems harder to justify than in classical physics: A quantum mechanical $N$-particle system possesses a wave function $\psi(\mathbf{x}_1,...\mathbf{x}_N,t)$ where $\mathbf{x}_1,...,\mathbf{x}_N$ are the particle coordinates. Suppose that a subsystem is formed by the particles with coordinates $q_1 = (\mathbf{x}_1,...,\mathbf{x}_{M})$. A simple way to associate a wave function with the subsystem exists if $\psi$ has the form
\begin{equation}
 \psi(q_1,q_2,t) = \varphi(q_1,t) \phi(q_2,t),
 \label{eq:entangledpsi}
\end{equation}
where $q_2 = (\mathbf{x}_{M+1}, ..., \mathbf{x}_N)$. Then $\varphi$ can be regarded as the wave function of the subsystem. However, the presence of entanglement exactly means that $\psi$ cannot be written in the form \eqref{eq:entangledpsi}. This leads us to the question: Which possibilities does quantum physics provide to describe the subsystem?\\
In open system quantum theory, a non-autonomous subsystem description is achieved by the reduced density matrix
\begin{equation}
 W_{\rm red}(q_1,q_1',t) = \int dq_2 \, \psi(q_1,q_2,t) \psi^*(q_1',q_2,t),
 \label{eq:wred}
\end{equation}
where the environment is ``traced out'', i.e. its actual state is ignored. The description is non-autonomous because the time evolution for $W_{\rm red}$ is not given by a closed equation. In contrast, the association of the subsystem with a subsystem's wave function usually involves a preparation procedure like an ideal measurement which we informally describe by
\begin{align}
 \psi(q_1,q_2,t=0) ~\overset{\rm Schr\ddot{o}dinger~evolution}{\longmapsto}~\sum_{\alpha} \varphi_\alpha(q_1,t) \phi_\alpha(q_2,t)~\overset{\rm collapse}{\longmapsto}~\varphi_{\alpha_0}(q_1,t) \phi_{\alpha_0}(q_2,t),
 \label{eq:idealmeasurement}
\end{align}
which leads to a projection onto a subsystem's  wave function -- a projection which depends on the wave function of the total system, the environment included. The projection procedure, also called the collapse of the wave function, is, however, not theoretically founded on the fundamental equations of quantum mechanics. This fact is also known as the measurement problem or objectification problem. In order to achieve a \textit{justification} of the quantum formalism, a subsystem analysis is needed \cite{ludwig,operators}.\\
In Bohmian mechanics (BM; see e.g. \cite{bh,qpwqp,duerr,holland}), the subsystem description is part of the theory and the objectification problem therefore does not even occur. For spin-less BM, one can define a \textit{conditional wave function} $\psi_{\rm cond}$ by plugging into the wave function of the larger system $\Psi(q_1,q_2,t)$ the actual Bohmian environment configuration $Q_2$ \cite{q_equilib}:
\begin{equation}
 \psi_{\rm cond}(q_1,t) = \frac{1}{\mathcal{N}}\Psi(q_1,Q_2(t),t),
 \label{eq:psicond}
\end{equation}
where $\mathcal{N}$ is a normalization factor such that $\|\psi_{\rm cond} \|=1$.\\
The evolution of the subsystem configuration $Q_1$ then only depends on the environment configuration via the conditional wave function. Besides, it is also possible to express conditional probabilities for the subsystem via $\psi_{\rm cond}$ only. Therefore, the conditional wave function is basic to the statistical analysis of BM \cite{q_equilib}.\\
However, the description of a subsystem by the conditional wave function is typically not autonomous. Nevertheless, in certain situations this can be attained and the conditional wave function then becomes effective \cite{q_equilib}, in the sense that even the implicit reference to the environment configuration (which is present in $\psi_{\rm cond}$) is lost. More precisely, we say that a subsystem has the \textit{effective wave function} $\psi_{\rm eff}$ (up to normalization) at time $t$ if the wave function $\Psi$ of the total system and the environment configuration $Q_2(t)$ satisfy
\begin{equation}
 \Psi(q_1,q_2,t) = \psi_{\rm eff}(q_1,t) \Phi(q_2,t) + \Psi^\perp(q_1,q_2,t),
 \label{eq:defeffwave}
\end{equation}
where $\Phi$ and $\Psi^\perp$ have \textit{macroscopically disjoint}\footnote{See  \cite{q_equilib} for details.} $q_2$-supports and $Q_2(t) \in {\rm supp}\, \Phi$.\\
The effective wave function reflects the idea of a quantum mechanical subsystem wave function that has been ``prepared'' by controlling the environment (e.g. experimental devices). $\psi_{\rm eff}$ persists as long as the Schrödinger evolution of the composite system does not destroy the effective product structure. If it exists, $\psi_{\rm eff} = \psi_{\rm cond}$.\\
For non-relativistic BM with spin, the conditional wave function does not exist anymore. Plugging the actual environment configuration into the wave function of the composite system yields a spinor-valued wave function which still contains all the spinor degrees of freedom of the environment: Denote the spin components of the wave function of the composite system by $\Psi^{s_1 s_2}$. Then naively generalizing eq. (\ref{eq:psicond}) yields
\begin{equation}
 \psi_{\rm cond}^{s_1 s_2}(q_1,t) = \frac{1}{\mathcal{N}}\Psi^{s_1 s_2}(q_1,Q_2(t),t),
 \label{eq:psicondspin}
\end{equation}
 which cannot be a wave function associated only with the subsystem, as the spin index of the environment is still present.\\
 The substitute is a \textit{conditional density matrix} \cite{dm} where the spinor degrees of freedom of the environment are traced out. Explicitly:
 \begin{equation}
   {W_{\rm cond}^{\rm nonrel}}_{s_1'}^{s_1}(q_1,q_1',t) = \frac{1}{\mathcal{N}} \sum_{s_2} \Psi^{s_1 s_2}(q_1,Q_2(t),t) \Psi^\D_{s_1's_2}(q_1,Q_2(t),t)
  \label{eq:wcond}
 \end{equation}
 where $(\cdot)^\D$ denotes the conjugate transposed.\\
 It turns out that the conditional density matrix functions in the same way  as the conditional wave function in the spin-less case \cite{dm}. Of course, there are also situations in which an effective wave function exists \cite{dm}. It is clear that if this is the case the conditional density matrix is pure and given by the effective wave function. The converse is less obvious. 

 In this paper we extend the subsystem description to relativistic BM with spin, namely to the hypersurface Bohm-Dirac (HBD) model \cite{hbd}. It turns out that the description leads to a non-trivial generalization of the non-relativistic conditional density matrix. We begin with a brief review of the HBD model and point out a Hilbert space adapted to the model. Next, we introduce the new conditional density matrix, analyze its properties and find a lifting to a density operator on said Hilbert space. Finally, we generalize the notion of the effective wave function and prove a lemma that clarifies its relation to the conditional density matrix.

\section{The hypersurface Bohm-Dirac model}
The hypersurface Bohm-Dirac model \cite{hbd} is a model of $N$ entangled, non-interacting\footnote{A fully consistent relativistic interacting quantum theory does not exist. However, we expect that the form of the subsystem description does not change in the presence of interaction.} Dirac (point) particles with actual world lines $X_k \in \mathbb{M},~k=1,...,N,$ where $\mathbb{M}$ is Minkowski spacetime with metric $g = {\rm diag}(1,-1,-1,-1)$. It uses a multi-time wave function \cite{dirac_32,dfp,tomonaga}
\begin{equation}
 \psi :~ \underbrace{\mathbb{M} \times \cdots \times \mathbb{M}}_{N~{\rm times}} ~~\longrightarrow ~~ (\mathbb{C}^{4})^{\otimes N}, ~~~~~
	 (x_1, ..., x_N) ~~\longmapsto ~~ \psi(x_1,..., x_N)
 \label{eq:multi-timewavefn}
\end{equation}
which is supposed to satisfy a multi-time system of non-interacting Dirac equations:
\begin{equation}
 \left[ i \gamma_k^{\mu_k} \partial_{k,\mu_k}  - m_k \right] \psi(x_1, ..., x_N) ~=~ 0, ~~~k=1,...,N, 
 \label{eq:multitimedirac}
\end{equation}
where $\partial_{k,\mu_k} = \frac{\partial }{\partial x_k^{\mu_k}}$, $\mu_k$ is the four-index of the $k$-th particle, $m_k$ stands for its mass and
\begin{equation}
 \gamma_k^{\mu_k} = 1 \otimes \cdots  \otimes 1 \underbrace{\otimes \gamma^{\mu_k} \otimes}_{k{\rm -th~ place}} 1 \otimes \cdots \otimes 1. 
 \label{eq:gamma}
\end{equation}
is the $\mu_k$-th Dirac gamma matrix acting on the spin index of the $k$-th particle.\\
As a consequence of eq. (\ref{eq:multitimedirac}) and its adjoint, the conserved tensor current reads:
\begin{equation}
 j^{\mu_1 \cdots \mu_N} = \overline{\psi} \gamma_{1}^{\mu_1} \cdots \gamma_{N}^{\mu_N} \psi,
\label{eq:jfree}
\end{equation}
where $\partial_{k,\mu_k} j^{\mu_1 \cdots \mu_k \cdots \mu_N} = 0,~k=1,...,N$. $\overline{\psi}$ stands for $\psi^\D \gamma_1^0 \cdots \gamma_N^0$.\\
The other crucial ingredient besides the multi-time wave function is a foliation $\mathcal{F}$ of spacetime into space-like hypersurfaces $\Sigma$, called ``leaves'' of $\mathcal{F}$. The foliation may be thought of as being generated by the wave function itself, thus rendering the theory relativistic (see \cite{rel_bm}  for a critical discussion of this idea).
Given such a foliation, the formulation of the relativistic guidance equation for the world lines is straightforward:
\begin{equation}
 \frac{d X_k^{\mu_k}(s)}{d s} \propto \left. j^{\mu_1 \cdots \mu_k \cdots \mu_N} (x_1, ..., x_N) \prod_{j\neq k} n_{\mu_j}(x_j) \right|_{x_i = X_i(\Sigma),~i=1,...,N}.
\label{eq:relguid}
\end{equation}
Here, $n$ denotes the time-like future-oriented unit normal vector field associated with $\mathcal{F}$ and $X_i(\Sigma)$ the intersection point of the $i$-th world line with $\Sigma \in \mathcal{F}$, and $X_k(s) \in \Sigma$. The proportionality sign expresses that the tangent vector $\dot{X}_k(\Sigma)$ should be parallel to the rhs (which is also a vector). This geometrical formulation implies the arbitrariness of the joint parametrization of the world lines via $s$.\\
The statistical import of the HBD model was analyzed in \cite{hbd}:
\begin{equation}
  \rho \equiv \overline{\psi} \, \gamma_1 \cdot n(x_1) \cdots \gamma_N \cdot n(x_N) \, \psi
  \label{eq:rho}
\end{equation}
obeys the continuity equation for curved surfaces and thus is an equivariant density on the leaves of the foliation, generalizing the well-known $|\psi|^2$-distribution.
Therefore, the crossing probability of $\Sigma \in \mathcal{F}$ is given by:
\begin{equation}
 {\rm Prob}\left({\rm particle}~i~{\rm crosses}~\Sigma~{\rm in}~d \sigma_i,~i=1,...,N\right) = \rho(x_1, ..., x_N) d \sigma_1 \cdots d \sigma_N,
 \label{eq:crossingprob}
\end{equation}
where $d \sigma_i$ denotes both an infinitesimal area on $\Sigma$ around $x_i$ as well as its 3-volume.\\
Note that the above formula for the crossing probability is only valid for hypersurfaces $\Sigma \in \mathcal{F}$. The HBD model so far does not make any statistical statements for hypersurfaces not belonging to the foliation. That this fact need not be in conflict with the usual quantum formalism has been remarked in \cite{hbd}.

\section{Mathematical interlude}

\subsection{Notation}
For the rest of the paper, we consider a HBD system $S$ (i.e. a set of particles) composed of two parts: $S_1$, the subsystem of interest and $S_2$, the environment. Schematically, we write this as $S = S_1 \cup S_2$. The number of particles splits according to $N = N_1 + N_2$. The wave function of $S$ is denoted by $\Psi$. We refer to its spin components as $\Psi^{s_1 s_2}$ and to the partial trace over these spin components as ${\rm tr}_{\mathbb{C}^{k_i}}$ where $k_i = 4^{N_i}$, $i = 1,2$. If $\Psi = \psi_1 \otimes \psi_2$, then $\overline{\psi}_1 = \psi_1^\D \gamma_1^0 \cdots \gamma_{N_1}^0$ and $\overline{\psi}_2 = \psi_2^\D \gamma_{N_1 + 1}^0 \cdots \gamma_{N}^0$ where these gamma-matrices now act on $\mathbb{C}^{k_1}$ and $\mathbb{C}^{k_2}$, respectively.  Consider a space-like hypersurface $\Sigma \in \mathcal{F}$ (we will use the symbol $\mathcal{S}$ for arbitrary space-like hypersurfaces). The configuration obtained by intersecting the world lines of the particles with $\Sigma$ is denoted by $Q(\
Sigma) = (Q_1(\Sigma),Q_2(\Sigma)) \in \Sigma^N$. A generic configuration spacetime variable is called $q = (q_1,q_2) = (x_1, ..., x_{N_1}, x_{N_1+ 1}, ..., x_N) \in \mathbb{M}^N$. The complex conjugate of $z \in \mathbb{C}$ is denoted by $z^*$.

\subsection{A relativistic Hilbert space} \label{sec:hilbertspace}
The appearance of the tensor current in the formula for the crossing probability suggests the definition of $N$-particle Hilbert spaces associated with space-like hypersurfaces $\mathcal{S}$:
\begin{equation}
  \mathcal{H}_{\mathcal{S}}^{(N)} = \left\{ \left. \chi : \mathcal{S}^N \rightarrow (\mathbb{C}^4)^{\otimes N} \right| \langle \chi, \chi \rangle_{\mathcal{S}}^{(N)} < \infty\right\},
  \label{eq:hilbertspace}
 \end{equation}
 with scalar product
 \begin{equation}
  \langle \psi, \varphi \rangle_{\mathcal{S}}^{(N)} = \int_{\mathcal{S}} d \sigma_{1,\mu_1} \cdots \int_{\mathcal{S}} d \sigma_{N,\mu_{N}} \overline{\psi} ~ \gamma_1^{\mu_1} \cdots \gamma_{N}^{\mu_{N}} \varphi.
  \label{eq:scalarprod}
 \end{equation}
 Here, $d \sigma_{k,\mu_k}$ stands for the product of $d \sigma_k$, the infinitesimal 3-volume element on $\mathcal{S}$, with $n_{\mu_k}$, the future-directed unit normal covector field at $\mathcal{S}$. The index $k$ refers to the fact that the integration is performed with respect to the variables of the $k$-th particle.\\
 The multi-time system of Dirac equations provides a multi-time wave function $\psi(q),~q \in \mathbb{M}^N$ on the whole of configuration spacetime. To have a wave function also on configurations $q \in \mathcal{S}^N$ where $\mathcal{S} \notin \mathcal{F}$ may be important for the case that the wave function is used to generate $\mathcal{F}$. This can be achieved already if the system of multi-time equations is only defined on $\mathscr{S} = \{ (x_1, ..., x_N) \in \mathbb{M}^N | (x_i - x_j)^2 < 0~\forall i \neq j\}$, the set of space-like configurations. The scenario of multi-time wave functions on $\mathscr{S}$ is mathematically analyzed in \cite{nogo_potentials,qftmultitime}.\\
 A wave function in $\mathcal{H}_{\mathcal{S}}^{(N)}$ can be obtained by restricting a multi-time wave function to configurations $q \in \mathcal{S}^N$. Schematically: $\psi_{|_\mathcal{S}}(q),~q \in \mathcal{S}^N$. The $\mathcal{H}_{\mathcal{S}}^{(N)}$ can be viewed as spaces of initial data for multi-time wave functions. The multi-time system of Dirac equations (\ref{eq:multitimedirac}) then yields a unitary evolution $U_{\mathcal{S} \rightarrow \mathcal{S}'}: \mathcal{H}_{\mathcal{S}}^{(N)} \rightarrow \mathcal{H}_{\mathcal{S}'}^{(N)}$ between the Hilbert spaces $\mathcal{H}_{\mathcal{S}}^{(N)}$ and $\mathcal{H}_{\mathcal{S}'}^{(N)}$ associated with different space-like hypersurfaces $\mathcal{S}$ and $\mathcal{S}'$.\footnote{A time evolution between curved surfaces was first considered by Tomonga in a QFT context \cite{tomonaga}.} We have: $U_{\mathcal{S} \rightarrow \mathcal{S}'} \psi_{|_\mathcal{S}} = \psi_{|_{\mathcal{S}'}}$. In particular, because of the uniqueness of solutions of eq. (\ref{eq:multitimedirac}) (cf. \cite{qftmultitime}): $\psi_{|_\mathcal{S}}(q) =  \psi_{|_{\mathcal{S}'}} (q) = \psi(q)$ for $q \in \mathcal{S} \cap \mathcal{S}'$.\\ 
 To see the unitarity of $U_{\mathcal{S} \rightarrow \mathcal{S}'}$, consider solutions $\psi, \varphi$ of eq. (\ref{eq:multitimedirac}) with compact support on any space-like hypersurface with respect to each of the coordinates $x_i$. Then the generalized tensor current
   \begin{equation}
  j^{\mu_1 \cdots \mu_N}[\psi, \varphi]  ~:=~ \overline{\psi} \gamma_1^{\mu_1} \cdots \gamma_N^{\mu_N} \varphi
  \label{eq:j2}
 \end{equation}
  is conserved, i.e. $\partial_{\mu_k} \overline{\psi} \gamma_1^{\mu_1} \cdots \gamma_k^{\mu_k} \cdots \gamma_N^{\mu_N} \varphi = 0,~k = 1, ..., N$.
The relation to the tensor current of eq. (\ref{eq:jfree}) is $j^{\mu_1 \cdots \mu_N} = j^{\mu_1 \cdots \mu_N}[\psi, \psi]$.\\
The scalar product of $\psi_{|_\mathcal{S}}$ and $\varphi_{|_\mathcal{S}}$ can be expressed as:
 \begin{equation}
   \langle \psi_{|_\mathcal{S}}, \varphi_{|_\mathcal{S}} \rangle_{\mathcal{S}}^{(N)} = \int_{\mathcal{S}} d \sigma_{1,\mu_1} \cdots \int_{\mathcal{S}} d \sigma_{N,\mu_{N}} \, j^{\mu_1 \cdots \mu_N}[\psi, \varphi].
  \label{eq:scalarprodj}
 \end{equation}
 The continuity equations that $j^{\mu_1 \cdots \mu_N}[\psi,\varphi]$ fulfils allow us to make use of the divergence theorem for a closed surface $\mathcal{S} \cup M \cup \mathcal{S}'$. Letting $M$ go to space-like infinity, the boundary terms drop out. This has to be done for each of the hypersurface integrals. Thus:
  \begin{align}
   \int_{\mathcal{S}} d \sigma_{1,\mu_1} \cdots \int_{\mathcal{S}} d \sigma_{N,\mu_{N}} \, j^{\mu_1 \cdots \mu_N}[\psi, \varphi] ~&=~ \int_{\mathcal{S}'} d \sigma_{1,\mu_1} \cdots \int_{\mathcal{S}'} d \sigma_{N,\mu_{N}} \, j^{\mu_1 \cdots \mu_N}[\psi, \varphi] \nonumber\\
   \Leftrightarrow ~ \langle \psi_{|_\mathcal{S}}, \varphi_{|_\mathcal{S}} \rangle_{\mathcal{S}}^{(N)} ~=~ \langle \psi_{|_{\mathcal{S}'}}, \varphi_{|_{\mathcal{S}'}} \rangle_{\mathcal{S}'}^{(N)} ~&=~ \langle  U_{\mathcal{S} \rightarrow \mathcal{S}'} \psi_{|_\mathcal{S}},U_{\mathcal{S} \rightarrow \mathcal{S}'} \varphi_{|_\mathcal{S}} \rangle_{\mathcal{S}'}^{(N)}.
  \label{eq:scalarprodhyperindpt2}
 \end{align}
 The scalar product (\ref{eq:scalarprod}) has a clear physical meaning for $\mathcal{S} = \Sigma,~\mathcal{S}' = \Sigma';~\Sigma,\Sigma' \in \mathcal{F}$:
 Then the norm of a wave function $\| \psi \|_\Sigma^{(N)} = \sqrt{\langle \psi ,\psi \rangle_{\Sigma}^{(N)}}$ gives unity by eq. (\ref{eq:crossingprob}), irrespective of $\Sigma$ due to eq. (\ref{eq:scalarprodhyperindpt2}). If furthermore $A \subset \Sigma^N$, $\langle \psi ,1_A \, \psi \rangle_{\Sigma}^{(N)}$ is the probability for  $Q(\Sigma) \in A$. Here, $1_A$ denotes the characteristic function of $A$.\\
 In the literature on the one-particle Dirac equation (see e.g. \cite{thaller}), one uses $L^2(\mathbb{R}^3) \otimes \mathbb{C}^4$ with scalar product $\langle \psi, \varphi \rangle = \int d^3 x \, \psi^\D \varphi$. The construction for $\mathcal{H}_{\mathcal{S}}^{(1)}$ reduces to this formula in the case of flat hypersurfaces, i.e. equal-time surfaces in a distinguished Lorentz frame. Because of this fact and the last point, we regard the use of $\mathcal{H}_{\mathcal{S}}^{(N)}$ instead of $L^2(\mathbb{R}^{3N}) \otimes (\mathbb{C}^4)^{\otimes N}$ as natural.

\section{Results}

\subsection{Conditional density matrix}
We first aim at a subsystem description of $S_1$ by a conditional density matrix.
To begin with, we rewrite eq. (\ref{eq:relguid}), observing (\ref{eq:jfree}), for the world lines of particles in $S_1$ by applying the identity
\begin{equation}
 v^\D w = \left( v_1^*, ..., v_k^* \right) \begin{pmatrix} w_1\\ \vdots\\ w_k \end{pmatrix} = v_1^* w_1 + ... + v_N^* w_N = {\rm tr}_{\mathbb{C}^k} \begin{pmatrix} v_1^* w_1 & ~ & *\\ ~ & \ddots & ~\\ * & ~ & v_N^* w_N \end{pmatrix} = {\rm tr}_{\mathbb{C}^k} (w v^\D)
 \label{eq:traceidentity}
\end{equation}
to the rhs of eq. (\ref{eq:relguid}) with $v^\D = \overline{\Psi}(q) \gamma_1^{\mu_1} ... \gamma_N^{\mu_N} \, \prod_{j\neq k} n_{\mu_j}(x_j)$ and $w = \Psi(q)$ for fixed $q$. This yields:
\begin{equation}
	\frac{d X_k^{\mu_k}(s)}{d s} ~\propto~ {\rm tr}_{\mathbb{C}^k} \, \left[ \Psi \overline{\Psi} \gamma_1^{\mu_1} \cdots \gamma_k^{\mu_k} \cdots \gamma_N^{\mu_N} \, \prod_{j\neq k} n_{\mu_j}(x_j) \right]_{q = Q(\Sigma)}.
\label{eq:derivwcondrel1}
\end{equation}
 Next, we split up the trace according to ${\rm tr}_{\mathbb{C}^{k}} \equiv {\rm tr}_{\mathbb{C}^{k_1}} \, {\rm tr}_{\mathbb{C}^{k_2}}$ and noting that the $\gamma$-matrices in eq. (\ref{eq:derivwcondrel1}) commute we obtain after rearranging:
\begin{align}
 \frac{d X_k^{\mu_k}(s)}{d s} &~\propto ~{\rm tr}_{\mathbb{C}^{k_1}} \, \left\{ {\rm tr}_{\mathbb{C}^{k_2}} \, \left[ \Psi \Psi^\D \prod_{j \in S_2} \gamma_j^0 \gamma_j^{\mu_j} n_{\mu_j}(x_j) \right]_{q_2 = Q_2(\Sigma)} \gamma_k^0 \gamma_k^{\mu_k} \prod_{j\neq k, \, j \in S_1} \gamma_j^0 \gamma_j \cdot n(x_j)  \right\}_{q_1 = Q_1(\Sigma)}.
\label{eq:derivwcondrel2}
\end{align}
In slight abuse of notation, in eq. (\ref{eq:derivwcondrel2}) we use the same symbols for the gamma matrices as before although now $\gamma_j^{\mu_j},~j=1,...,N_1,$ act on $\mathbb{C}^{k_1}$ instead of $\mathbb{C}^{k}$. The separation of variables associated with $S_1$ and $S_2$ in eq. \eqref{eq:derivwcondrel2} leads to a rewriting of the HBD guidance law of the desired form:
\begin{equation}
 \frac{d X_k^{\mu_k}(s)}{d s} ~\propto ~ {\rm tr}_{\mathbb{C}^{k_1}} \, \left\{ W_{\rm cond}(q_1, q_1') ~ \gamma_k^0 \gamma_k^{\mu_k} \prod_{j\neq k, \, j \in S_1} \gamma_j^0 \gamma_j \cdot n(x_j) \right\}_{q_1 = q_1' = Q_1(\Sigma)},
 \label{eq:derivwcondrel}
\end{equation}
where
\begin{align}
	{W_{\rm cond}}^{s_1}_{s_1'}(q_1, q_1') ~&:=~ \frac{1}{\mathcal{N}} \sum_{s_2} \Psi^{s_1 s_2}(q_1, Q_2(\Sigma)) \left[ \Psi^\D (q_1', Q_2(\Sigma)) \prod_{j \in S_2} \gamma_j^0 \gamma_j \cdot n(X_j(\Sigma)) \right]_{s_1' s_2}
\label{eq:wcondrel}
\end{align}
define the components of the \textit{conditional density matrix} for $S_1$ and
\begin{equation}
 \mathcal{N} := \int_{\Sigma} d \sigma_{1} \cdots \int_{\Sigma} d \sigma_{N_1} {\rm tr}_{\mathbb{C}^{k}} \left[ \Psi(q_1,Q_2(\Sigma)) \Psi^\D(q_1,Q_2(\Sigma)) \prod_{j \in S_2} \gamma_j^0 \gamma_j \cdot n(X_j(\Sigma)) \prod_{j \in S_1} \gamma_j^0 \gamma_j \cdot n(x_j)\right]
\label{eq:wcondrelnormal}
\end{equation}
is the appropriate normalization factor. Note that $\mathcal{N}$ is actually independent of the choice of the space-like hypersurface in the domain of integration. To see this, we write:
\begin{align}
  \mathcal{N} ~&=~ \int_{\Sigma} d \sigma_{1} \cdots \int_{\Sigma} d \sigma_{N_1} \Psi^\D(q_1,Q_2(\Sigma)) \prod_{j \in S_2} \gamma_j^0 \gamma_j \cdot n(X_j(\Sigma)) \prod_{j \in S_1} \gamma_j^0 \gamma_j \cdot n(x_j) \Psi(q_1,Q_2(\Sigma)) \nonumber\\
 &=~ \int_{\Sigma} d \sigma_{1,\mu_1} \cdots \int_{\Sigma} d \sigma_{N_1,\mu_{N_1}} \, \left. j^{\mu_1 \cdots \mu_{N_1} \mu_{N_1+1} \cdots \mu_N}[\Psi,\Psi] \right|_{q_2 = Q_2(\Sigma)} \prod_{j \in S_2} n_{\mu_j}(X_j(\Sigma)),
 \label{eq:wcondrelnormal2}
\end{align}
which can be shown to be independent of the hypersurface in the domain of integration following the steps leading from eq. (\ref{eq:scalarprodj}) to eq. (\ref{eq:scalarprodhyperindpt2}).\\
Also note that for a flat foliation, i.e. $n(x) = (1,0,0,0)~\forall x$ in a certain frame, our definition of $W_{\rm cond}$ coincides with the one for non-relativistic BM with spin in that frame (cf. eq. (\ref{eq:wcond}) and since $(\gamma_k^0)^2 = 1 \, \forall k$).\\

\noindent The physical significance of $W_{\rm cond}$ is based on the dynamical role of $W_{\rm cond}$ as well as its role in the statistical analysis. Eq. (\ref{eq:derivwcondrel}) establishes the dynamical role of $W_{\rm cond}$. In order to analyze the statistical meaning of $W_{\rm cond}$, we start from the crossing probability of $\Sigma \in \mathcal{F}$ of the HBD model (eq. (\ref{eq:crossingprob})):
 \begin{equation}
  {\rm Prob}\left({\rm particle}~i~{\rm crosses}~\Sigma~{\rm in}~d \sigma_i,~i=1,...,N\right) = \rho(x_1, ..., x_N) d \sigma_1 \cdots d \sigma_N.
  \label{eq:crossingprob2}
 \end{equation}
 Next, we condition:
 \begin{equation}
  {\rm Prob}\left({\rm particle}~i~{\rm crosses}~\Sigma~{\rm in}~d \sigma_i,~i=1,...,N_1 | Q_2(\Sigma)\right) = \frac{\rho(x_1, ..., x_{N_1},Q_2(\Sigma)) d \sigma_1 \cdots d \sigma_{N_1}}{\int_{\Sigma} d \sigma_1 \cdots \int_{\Sigma} d \sigma_{N_1} \rho(x_1, ..., x_{N_1},Q_2(\Sigma))}
  \label{eq:crossingprob3}
 \end{equation}
and comparing eq. (\ref{eq:jfree}) with
 \begin{equation}
  \rho(x_1, ..., x_N) = \overline{\Psi}(q) \gamma_{1}^{\mu_1}\cdots \gamma_{N}^{\mu_N} \Psi(q) \prod_{j=1}^N n_{\mu_j}(x_j) ~~~\left( = j^{\mu_1 \cdots \mu_N} \prod_{j=1}^N n_{\mu_j}(x_j) \right),
 \end{equation}
 we repeat the same steps leading from eq. (\ref{eq:relguid}) to eqs. (\ref{eq:derivwcondrel1})-(\ref{eq:derivwcondrel}) to obtain:
 \begin{align}
  & {\rm Prob}\left({\rm particle}~i~{\rm crosses}~\Sigma~{\rm in}~d \sigma_i,~i=1,...,N_1 | Q_2(\Sigma)\right) \nonumber\\
  &~~~ =~ {\rm tr}_{\mathbb{C}^{k_1}} \left[ W_{\rm cond}(q_1,q_1) \prod_{j \in S_1} \gamma_j^0 \gamma_j \cdot n(x_j) \right] d \sigma_1 \cdots d \sigma_{N_1}. 
 \label{eq:wcondprobrel}
 \end{align}
 Eqs. (\ref{eq:crossingprob3}) and (\ref{eq:wcondprobrel}) also explain why the normalization (\ref{eq:wcondrelnormal}) of $W_{\rm cond}$ is appropriate: $\mathcal{N} = \int_{\Sigma} d \sigma_1 \cdots \int_{\Sigma} d \sigma_{N_1} \rho(x_1, ..., x_{N_1},Q_2(\Sigma))$.\\
 Eq. (\ref{eq:wcondprobrel}) allows us to calculate expectation values, e.g. of the distribution of the subsystem configuration. Let $\hat{Q}_1$ denote the multiplication operator with $q_1$ on $\mathcal{H}_\Sigma^{(N_1)}$ for $\Sigma \in \mathcal{F}$; then its expectation value for a ``state'' characterized by $W_{\rm cond}$ is given by:
 \begin{equation}
  \langle \hat{Q}_1 \rangle_{W_{\rm cond}} = \int_{\Sigma} d \sigma_{1} \cdots \int_{\Sigma} d \sigma_{N_1} ~q_1 \, {\rm tr}_{\mathbb{C}^{k_1}} \left[ W_{\rm cond}(q_1,q_1) \prod_{j \in S_1} \gamma_j^0 \gamma_j\cdot n(x_j) \right].
  \label{eq:expectationq1}
 \end{equation}

 \subsection{Conditional density operator}
 We introduce an operator $\hat{W}_{\rm cond}$ on $\mathcal{H}_{\mathcal{S}}^{(N_1)}$ such that for $\mathcal{S} = \Sigma \in \mathcal{F}$, eq. (\ref{eq:expectationq1}) can be rewritten as the trace ${\rm tr}(\hat{W}_{\rm cond} \, \hat{Q_1})$.\\
 For this purpose, we define bras $\langle q_1,s_1 |$ and kets $| q_1',s_1' \rangle$ with $\langle q_1,s_1 | q_1',s_1' \rangle = \delta^{(3 N_1)}(q_1-q_1') \delta_{s_1 s_1'}$ and $\hat{Q}_1 | q_1,s_1 \rangle = q_1 | q_1,s_1 \rangle$ where $\langle \cdot, \cdot \rangle$ is the scalar product (\ref{eq:scalarprod}) on $\mathcal{H}_{\mathcal{S}}^{(N_1)}$ (cf. (\ref{eq:hilbertspace})).\\
 Let
\begin{equation}
 (\gamma n)(q_i) := \prod_{j \in S_i} \gamma_j^0 \gamma_j \cdot n(x_j),~~~{\rm where}~~~(\gamma n)^\D(q_i) = (\gamma n)(q_i),~~~i = 1,2,
 \label{eq:gamman}
\end{equation}
  since $(\gamma_k^{\mu_k})^\D = \gamma_k^0 \gamma_k^{\mu_k} \gamma_k^0$.\\
  We now show that for every fixed $q_i$, $(\gamma n)(q_i),~i=1,2,$ is a positive matrix. To see this, we consider the quadratic forms $\Psi^\D \,(\gamma n)(q_i) \, \Psi,~i=1,2,$ and rewrite them using eqs. (\ref{eq:gamman}) and (\ref{eq:j2}):
 \begin{align}
  \Psi^\D \,(\gamma n)(q_1) \, \Psi ~&=~ j^{\mu_1 \cdots \mu_{N_1} 0 \cdots 0}[\Psi,\Psi] \prod_{j \in S_1} n_{\mu_j}(x_j),\nonumber\\
  \Psi^\D \,(\gamma n)(q_2) \, \Psi ~&=~ j^{0 \cdots 0 \, \mu_{N_1+1} \cdots \mu_N }[\Psi,\Psi] \prod_{j \in S_2} n_{\mu_j}(x_j),
  \label{eq:quadraticform}
 \end{align}
 which both are greater or equal to zero as the tensor current $j^{\mu_1 ... \mu_N}[\Psi,\Psi]$ is positive-definite.\\
 As a consequence of the positivity and the self-adjointness of $(\gamma n)(q_i)$, it is possible to define $\sqrt{(\gamma n)(q_i)}$.
 The relation of $ \langle q_1, s_1 | \varphi \rangle$ to the components $\varphi^{s_1}(q_1)$ of a wave function $\varphi$ in $\mathcal{H}_{\mathcal{S}}^{(N_1)}$ then is:
 \begin{equation}
  \langle q_1, s_1 | \varphi \rangle = \sum_{s_1'} \sqrt{(\gamma n)(q_1)}_{s_1'}^{s_1} \varphi^{s_1'}(q_1).
  \label{eq:relationcomponents}
 \end{equation}
 Using the abbreviation
 \begin{equation}
  \int_{\mathcal{S}^{N_1}} d^{3N_1} q_1 ~:=~ \int_{\mathcal{S}} d \sigma_1 \cdots \int_{\mathcal{S}} d \sigma_{N_1},
  \label{eq:intabbrev}
 \end{equation}
 we put:
 \begin{align}
  {\rm tr}(\hat{W}_{\rm cond} \, \hat{Q_1}) ~&=~ \sum_{s_1} \int_{\mathcal{S}^{N_1}} d^{3N_1} q_1 \, \langle q_1,s_1| \hat{W}_{\rm cond} \, \hat{Q}_1 | q_1, s_1 \rangle \nonumber\\
  &=~  \sum_{s_1} \int_{\mathcal{S}^{N_1}} d^{3N_1} q_1 ~ q_1 \, \langle q_1,s_1| \hat{W}_{\rm cond} | q_1, s_1 \rangle.
  \label{eq:traceconsideration}
 \end{align}
 Comparing eq. (\ref{eq:traceconsideration}) with eq. (\ref{eq:expectationq1}), we are led to define:
 \begin{equation}
  \langle q_1,s_1| \hat{W}_{\rm cond} | q_1', s_1' \rangle := \left( \sqrt{(\gamma n)(q_1)} \, W_{\rm cond}(q_1,q_1') \sqrt{(\gamma n)(q_1')} \right)_{s_1'}^{s_1}.
  \label{eq:defwhat1}
 \end{equation}
 The action of $\hat{W}_{\rm cond}$ on a vector $\varphi \in \mathcal{H}_{\mathcal{S}}^{(N_1)}$, expressed in components, is given by:
\begin{equation}
 (\hat{W}_{\rm cond} \, \varphi )^{s_1}(q_1) := \int_{\mathcal{S}^{N_1}} d^{3N_1} q_1' \sum_{s_1'} \left[ W_{\rm cond}(q_1,q_1') (\gamma n)(q_1') \right]_{s_1'}^{s_1} \varphi^{s_1'}(q_1').
 \label{eq:actionwcond}
\end{equation}
Note that eqs. (\ref{eq:relationcomponents}), (\ref{eq:defwhat1}) and (\ref{eq:actionwcond}) are chosen consistently, as one must have:
 \begin{align}
  &\left[ \sqrt{(\gamma n) (q_1)} \, \hat{W}_{\rm cond} \, \varphi \right]^{s_1}(q_1) ~\stackrel{\rm (\ref{eq:relationcomponents})}{=}~ \langle q_1, s_1 | \hat{W}_{\rm cond} \, | \varphi \rangle \nonumber\\
 &~~~~~=~ \int_{\mathcal{S}^{N_1}} d^{3N_1} q_1' \sum_{s_1'} \langle q_1, s_1 | \hat{W}_{\rm cond} \, | q_1', s_1' \rangle \langle q_1', s_1' | \varphi \rangle\nonumber\\
 &~~\stackrel{(\ref{eq:defwhat1}), (\ref{eq:relationcomponents})}{=}~   \int_{\mathcal{S}^{N_1}} d^{3N_1} q_1' \sum_{s_1'} \left( \sqrt{(\gamma n)(q_1)} \, W_{\rm cond}(q_1,q_1') \sqrt{(\gamma n)(q_1')} \right)_{s_1'}^{s_1} \left[ \sqrt{(\gamma n)(q_1')} \varphi(q_1')\right]^{s_1'} \nonumber\\
 &~~~~\stackrel{(\ref{eq:actionwcond})}{=}~ \left[ \sqrt{(\gamma n) (q_1)} \, \hat{W}_{\rm cond} \, \varphi \right]^{s_1}(q_1).
  \label{eq:consistencywcond}
 \end{align}
In fact, as one expects, $\hat{W}_{\rm cond}$ (as an operator on $\mathcal{H}_{\Sigma}^{(N_1)}$)\footnote{Note that the restriction to $\mathcal{H}_{\Sigma}^{(N_1)}$ instead of a general $\mathcal{H}_{\mathcal{S}}^{(N_1)}$ is necessary because of the use of $\langle q_1, s_1, Q_2(\Sigma), s_2 |$ on $\mathcal{H}_{\Sigma}^{(N)}$ (cf. eq. (\ref{eq:particaltracewcond})) which requires $(q_1,Q_2(\Sigma)) \in \Sigma^N$.} can equivalently be derived from the projector $| \Psi \rangle \langle \Psi |$ on $\mathcal{H}_{\Sigma}^{(N)}$ by a partial trace. For this purpose, we straightforwardly generalize eq. (\ref{eq:relationcomponents}) to $\mathcal{H}_{\Sigma}^{(N)}$:
\begin{equation}
  \langle q_1, s_1, q_2, s_2 | \Psi \rangle ~=~ \sum_{s_1' s_2'} \sqrt{(\gamma n)(q_1)}_{s_1'}^{s_1} \sqrt{(\gamma n)(q_2)}_{s_2'}^{s_2} \, \Psi^{s_1' s_2'}(q_1,q_2).
  \label{eq:relationcomponents2}
 \end{equation}
Then:
\begin{align}
 &\sum_{s_2} \langle q_1,s_1,Q_2(\Sigma),s_2| \Psi \rangle \langle \Psi | q_1', s_1', Q_2(\Sigma), s_2 \rangle\nonumber\\
 &~~~\stackrel{(\ref{eq:relationcomponents2})}{=}~ \sum_{s_2} \sum_{\tilde{s}_1,\tilde{s}_2} \sqrt{(\gamma n)(q_1)}_{\tilde{s}_1}^{s_1} \, \sqrt{(\gamma n)(Q_2(\Sigma))}_{\tilde{s}_2}^{s_2} \Psi^{\tilde{s}_1 \tilde{s}_2}(q_1,Q_2(\Sigma)) \nonumber\\
&~~~~~~~\times \sum_{\hat{s}_1, \hat{s}_2} {\Psi^\D}_{\hat{s}_1,\hat{s}_2}(q_1,Q_2(\Sigma)) \sqrt{(\gamma n)(q_1')}_{s_1'}^{\hat{s}_1} \, \sqrt{(\gamma n)(Q_2(\Sigma))}_{s_2}^{\hat{s}_2} \nonumber\\
&~~~\stackrel{(\ref{eq:wcondrel})}{=}~ \sum_{\tilde{s}_1,\hat{s}_1} \sqrt{(\gamma n)(q_1)}_{\tilde{s}_1}^{s_1} \, {W_{\rm cond}}_{\hat{s}_1}^{\tilde{s}_1}(q_1,q_1') \sqrt{(\gamma n)(q_1')}_{s_1'}^{\hat{s}_1} \nonumber\\
&~~~\stackrel{(\ref{eq:defwhat1})}{=}~\langle q_1,s_1| \hat{W}_{\rm cond} | q_1', s_1' \rangle.
 \label{eq:particaltracewcond}
\end{align}
As expected, we have the following

\paragraph{Lemma:} \textit{$\hat{W}_{\rm cond}$ is a density operator on $\mathcal{H}_{\mathcal{S}}^{(N_1)}$.}\\[0.2cm]
\noindent \textit{Proof:}\\[0.2cm]
 1. Consider $\mathcal{H}_{\mathcal{S}}^{(N_1)}$ for a general space-like hypersurface $\mathcal{S}$. Then $\hat{W}_{\rm cond}$ is self-adjoint on $\mathcal{H}_{\mathcal{S}}^{(N_1)}$.
 In the proof we make use of the property $W_{\rm cond}^\D(q_1,q_1') = W_{\rm cond}(q_1',q_1)$. To see this, consider:
 \begin{equation}
  W_{\rm cond}^\D(q_1,q_1') =  \left\{ {\rm tr}_{\mathbb{C}^{k_2}} \left[ \Psi(q_1,Q_2(\Sigma)) \Psi^\D(q_1',Q_2(\Sigma)) (\gamma n)(Q_2(\Sigma)) \right] \right\}^\D.
  \label{eq:wcondsa1}
 \end{equation}
 With the identity $({\rm tr}_{\mathbb{C}^{k_2}} A)^\D =  {\rm tr}_{\mathbb{C}^{k_2}} (A^\D)$ for the partial trace of matrices $A: \mathbb{C}^{k} \rightarrow \mathbb{C}^{k}$, eq. \eqref{eq:wcondsa1} becomes:
 \begin{equation}
  W_{\rm cond}^\D(q_1,q_1') =  {\rm tr}_{\mathbb{C}^{k_2}} \left[ (\gamma n)^\D(Q_2(\Sigma)) \Psi(q_1',Q_2(\Sigma)) \Psi^\D(q_1,Q_2(\Sigma)) \right].
  \label{eq:wcondsa2}
 \end{equation} 
Using the cyclic property of the partial trace, i.e. ${\rm tr}_{\mathbb{C}^{k_2}}( 1 \otimes B \, A ) = {\rm tr}_{\mathbb{C}^{k_2}}( A \, 1 \otimes B )$ where $B: \mathbb{C}^{k_2} \rightarrow \mathbb{C}^{k_2}$, and observing \eqref{eq:gamman} we finally obtain:
  \begin{equation}
  W_{\rm cond}^\D(q_1,q_1') ~=~  {\rm tr}_{\mathbb{C}^{k_2}} \left[ \Psi(q_1',Q_2(\Sigma)) \Psi^\D(q_1,Q_2(\Sigma)) (\gamma n)(Q_2(\Sigma)) \right] = W_{\rm cond}(q_1',q_1).
  \label{eq:wcondsa3}
 \end{equation}
 We proceed with the proof of the self-adjointness of $\hat{W}_{\rm cond}$:
\begin{align}
 \langle \psi, \hat{W}_{\rm cond} \,\varphi \rangle_{\mathcal{S}}^{(N_1)} &\stackrel{(\ref{eq:scalarprod}), (\ref{eq:actionwcond})}{=} \int_{\mathcal{S}^{N_1}} d^{3N_1} q_1 \int_{\mathcal{S}^{N_1}} d^{3N_1} q_1' \, \psi^\D(q_1) \, (\gamma n)(q_1) \, W_{\rm cond}(q_1,q_1') \, (\gamma n)(q_1') \, \varphi(q_1') \nonumber\\
 & \stackrel{(\ref{eq:gamman}), (\ref{eq:wcondsa3})}{=} \int_{\mathcal{S}^{N_1}} d^{3N_1} q_1 \int_{\mathcal{S}^{N_1}} d^{3N_1} q_1' \left[ W_{\rm cond}(q_1',q_1) \, (\gamma n)(q_1) \, \psi(q_1) \right]^\D (\gamma n)(q_1') \, \varphi(q_1')\nonumber\\
 &~~\stackrel{(\ref{eq:actionwcond})}{=}~\int_{\mathcal{S}^{N_1}} d^{3N_1} q_1' \left[ \hat{W}_{\rm cond} \,  \psi \right]^\D(q_1') \, (\gamma n)(q_1') \, \varphi(q_1')\nonumber\\
 &~~\stackrel{(\ref{eq:scalarprod})}{=}~\langle \hat{W}_{\rm cond} \, \psi,  \,\varphi \rangle_{\mathcal{S}}^{(N_1)}.
 \label{eq:wcondsym}
\end{align}
Thus, $\hat{W}_{\rm cond}$ is symmetric on $\mathcal{H}_{\mathcal{S}}^{(N_1)}$ and as a bounded operator it is therefore also self-adjoint.\\[0.2cm]
2. To show the positivity of $\hat{W}_{\rm cond}$, we start with
\begin{equation}
 \langle \varphi, \hat{W}_{\rm cond} \,\varphi \rangle_{\mathcal{S}}^{(N_1)} ~=~ \int_{\mathcal{S}^{N_1}} d^{3N_1} q_1 \int_{\mathcal{S}^{N_1}} d^{3N_1} q_1' \, \varphi^\D(q_1) \, (\gamma n)(q_1) \, W_{\rm cond}(q_1,q_1') \, (\gamma n)(q_1') \, \varphi(q_1')
 \label{eq:wcondpos}
\end{equation}
and plug in the explicit form of ${W_{\rm cond}}_{s_1'}^{s_1}(q_1,q_1')$ from eq. (\ref{eq:wcondrel}). Next, we simplify the expression for ${W_{\rm cond}}_{s_1'}^{s_1}(q_1,q_1')$ by a multi-time Lorentz transformation, i.e. a transformation of the form
 \begin{equation}
   \tilde{\Lambda} \equiv (\Lambda_1, ..., \Lambda_N):~\mathbb{M}^N \rightarrow \mathbb{M}^N,~~~(x_1, ..., x_N) \longmapsto (\Lambda_1 x_1, ..., \Lambda_N x_N),
 \label{eq:multilorentztrafo}
 \end{equation}
 where $\Lambda_i$ are the matrices forming the usual representation of the (connected part of the) Lorentz group on the $i$-th copy of $\mathbb{M}$. Under the action of these transformations, a multi-time wave function behaves as follows:
 \begin{equation}
   \psi(x_1, ..., x_N) ~~~ \stackrel{\tilde{\Lambda}}{\longmapsto} ~~~\psi'(x_1,...,x_N) ~= ~ S[\Lambda_1] \otimes \cdots \otimes S[\Lambda_N] \, \psi(\Lambda_1^{-1} x_1, ..., \Lambda_N^{-1} x_N),
 \label{eq:multi-timespinorrep}
 \end{equation}
 where $S[\Lambda_i]$ denote the matrices forming the usual spinor representation of the Lorentz group.\\
 Now, as for fixed $Q(\Sigma)$ all $n(X_j(\Sigma))$ are constant time-like future-oriented unit vectors, we may choose $\tilde{\Lambda}$ such that $n'(X_j(\Sigma)) \equiv (1,0,0,0),~j = 1,..., N$. Thus: $(\gamma n')(Q_2(\Sigma)) = 1$.\\
 Continuing with the hereby simplified eq. \eqref{eq:wcondpos} and dropping the Lorentz transformation primes for notational ease, we have:
\begin{align}
 \langle \varphi, \hat{W}_{\rm cond} \,\varphi \rangle_{\mathcal{S}}^{(N_1)} ~&=~ \frac{1}{\mathcal{N}} \int_{\mathcal{S}^{N_1}} d^{3N_1} q_1 \int_{\mathcal{S}^{N_1}} d^{3N_1} q_1' \sum_{s_1, s_1'} \left\{ \left[ \varphi^\D(q_1) \, (\gamma n)(q_1)\right]_{s_1} \right. \nonumber\\
&~~~~~\times ~\left. \sum_{s_2} \Psi^{s_1 s_2}(q_1, Q_2(\Sigma)) \Psi^\D_{s_1' s_2} (q_1', Q_2(\Sigma)) \left[ (\gamma n)(q_1') \, \varphi(q_1') \right]^{s_1'} \right\}\nonumber\\
 &=~ \frac{1}{\mathcal{N}} \sum_{s_2}  \left(\int_{\mathcal{S}^{N_1}} d^{3N_1} q_1 \sum_{s_1} \left[ \varphi^\D(q_1) \, (\gamma n)(q_1)\right]_{s_1} \Psi^{s_1 s_2}(q_1, Q_2(\Sigma)) \right) \nonumber\\
 &~~~~~~~~~~\times \left(\int_{\mathcal{S}^{N_1}} d^{3N_1} q_1' \sum_{s_1'} \Psi^\D_{s_1' s_2} (q_1', Q_2(\Sigma)) \left[ (\gamma n)(q_1') \, \varphi(q_1') \right]^{s_1'} \right) \nonumber\\
 &\equiv~ \sum_{s_2} c^{s_2} c_{s_2}^* ~\geq 0,
 \label{eq:wcondpos2}
\end{align}
where
\begin{equation}
 c^{s_2} ~=~ \frac{1}{\sqrt{\mathcal{N}}} \int_{\mathcal{S}^{N_1}} d^{3N_1} q_1 \sum_{s_1} \left[ \varphi^\D(q_1) \, (\gamma n)(q_1)\right]_{s_1} \Psi^{s_1 s_2}(q_1, Q_2(\Sigma)).
\label{eq:cs2}
\end{equation}
~\\
3. $\hat{W}_{\rm cond}$ is normalized. This straightforwardly follows from the fact that ${\rm tr} \, \hat{W}_{\rm cond}$ yields unity by eq. (\ref{eq:wcondprobrel}) when one applies the computational formula for the trace as used in eq. (\ref{eq:traceconsideration}) and makes use of eq. (\ref{eq:defwhat1}). $\Box$

\subsection{Effective wave function} \label{sec:effwave}
Because of the dependence on $Q_2(\Sigma)$, $W_{\rm cond}$ typically does not evolve autonomously, i.e. according to its own multi-time system of von Neumann equations. We now turn to an autonomous subsystem description in terms of wave functions. For this purpose, we extend the definition of the effective wave function (cf. eq. (\ref{eq:defeffwave})) to the HBD model. Assume that there exists a hypersurface $\Sigma \in \mathcal{F}$ such that $\Psi^{s_1 s_2}(q_1, q_2)$ and the actual configuration $Q(\Sigma) = (Q_1,Q_2)(\Sigma)$ on that hypersurface satisfy
\begin{equation}
    \Psi^{s_1 s_2}(q_1,q_2) = \psi_1^{s_1}(q_1) \psi_2^{s_2}(q_2) + (\Psi^\perp)^{s_1 s_2}(q_1,q_2)~~~\forall q = (q_1, q_2) \in \Sigma^N,
    \label{eq:effwavemulti-time}
\end{equation}
with $\psi_2$ and $\Psi^\perp$ possessing \textit{macroscopically disjoint} $q_2$-supports and $Q_2(\Sigma) \in {\rm supp} \, \psi_2$. Then for $q_1 \in \Sigma^{N_1}$ system $S_1$ is said to have \textit{effective wave function}
\begin{equation}
  \psi_{\rm eff}^{s_1}(q_1) = \frac{\psi_1^{s_1}(q_1)}{\| \psi_1 \|_\Sigma^{(N_1)}}.
  \label{eq:defeffwavemulti-time}
\end{equation}
It has the following properties:\\[0.2cm]
 1. If $\Sigma$ is an equal-time hypersurface of a Lorentz frame, then our definition agrees with the non-relativistic one in that frame.\\[0.2cm]
 2. Inserting the product structure $\Psi(q_1,Q_2(\Sigma)) = \psi_1(q_1) \otimes \psi_2(Q_2(\Sigma))$ resulting from eq. (\ref{eq:effwavemulti-time}) into the relativistic guidance equation (\ref{eq:relguid}) for a particle in $S_1$, we obtain:
 \begin{align}
  \frac{d X_k^{\mu_k}(s)}{d s} &\propto~ \left. \overline{\psi}_1(q_1) \otimes \overline{\psi}_2(Q_2(\Sigma))\, \gamma_k^{\mu_k}\prod_{j\neq k, j\in S_1} \gamma_j \cdot n(x_j) ~ (\gamma n)(Q_2(\Sigma)) \, \psi_1(q_1) \otimes \psi_2(Q_2(\Sigma))  \right|_{q_1 = Q_1(\Sigma)} \nonumber\\
 &\propto~ \left. \overline{\psi}_{\rm eff}(q_1) \gamma_1^{\mu_1} \cdots \gamma_k^{\mu_k} \cdots \gamma_{N_1}^{\mu_{N_1}} \psi_{\rm eff}(q_1)  \prod_{j\neq k, j \in S_1} n_{\mu_j}(x_j) \right|_{q_1 = Q_1(\Sigma)},
  \label{eq:effwaverelguid}
 \end{align}
 where we absorbed the factor $\| \psi_1 \|_\Sigma^{(N_1)} \,  \overline{\psi}_2(Q_2(\Sigma)) \,(\gamma n)(Q_2(\Sigma))\,  \psi_2(Q_2(\Sigma)) \geq 0$ into the proportionality. One obtains the same result as in eq. (\ref{eq:effwaverelguid}) if one starts with a pure\footnote{We apply the notion \textit{pure} to the  the matrix $W_{\rm cond}$, in the sense that it can be written as $W_{\rm cond}(q_1,q_1') = \psi(q_1) \psi^\D(q_1')$ for some wave function $\psi$. Otherwise, we call it \textit{mixed}.} $W_{\rm cond}$ in eq. (\ref{eq:derivwcondrel}).\\[0.2cm]
 3. We now come to the expression of conditional probabilities, starting from the conditional version of the crossing probability (cf. eq. (\ref{eq:crossingprob3})):
  \begin{equation}
    {\rm Prob}\left({\rm particle}~i~{\rm crosses}~\Sigma~{\rm in}~d \sigma_i,~i=1,...,N_1 | Q_2(\Sigma)\right) = \frac{\rho(q_1,Q_2(\Sigma)) d \sigma_1 \cdots d \sigma_{N_1}}{\int_{\Sigma^{N_1}} d q_1^{3N_1} \, \rho(q_1,Q_2(\Sigma))}.
    \label{eq:crossingprob4}
  \end{equation}
 Into this equation we plug in the HBD density
 \begin{equation}
  \rho(q) = \overline{\Psi}(q) (\gamma n)(q_1) \, (\gamma n)(q_2) \Psi(q)
 \end{equation}
 in the particular situation given by eq. (\ref{eq:effwavemulti-time}), using that then $\Psi(q_1,Q_2(\Sigma)) = \psi_1(q_1)\otimes \psi_2(Q_2(\Sigma))$. This yields:
 \begin{align}
   & {\rm Prob}\left({\rm particle}~i~{\rm crosses}~\Sigma~{\rm in}~d \sigma_i,~i=1,...,N_1 | Q_2(\Sigma)\right)\nonumber\\
   ~~&=~ \frac{\overline{\psi}_1(q_1) \otimes \overline{\psi}_2(Q_2(\Sigma)) \, (\gamma n)(q_1) \, (\gamma n)(Q_2(\Sigma)) \, \psi_1(q_1) \otimes \psi_2(Q_2(\Sigma)) \, d \sigma_1 \cdots d \sigma_{N_1}}{\int_{\Sigma^{N_1}} d q_1^{3N_1} \, \overline{\psi}_1(q_1) \otimes \overline{\psi}_2(Q_2(\Sigma)) \, (\gamma n)(q_1) \, (\gamma n)((Q_2(\Sigma)))  \, \psi_1(q_1) \otimes \psi_2(Q_2(\Sigma))}\nonumber\\
   ~~&=~ \frac{\overline{\psi}_1(q_1) \, (\gamma n)(q_1) \, \psi_1(q_1) \, d \sigma_1 \cdots d \sigma_{N_1}}{\int_{\Sigma^{N_1}} d q_1^{3N_1} \, \overline{\psi}_1(q_1) \, (\gamma n)(q_1) \, \psi_1(q_1)}\nonumber\\
   ~~&=~ \overline{\psi}_{\rm eff}(q_1) \, (\gamma n)(q_1) \, \psi_{\rm eff}(q_1) \, d \sigma_1 \cdots d \sigma_{N_1}.
   \label{eq:probeffwavemulti-time}
 \end{align}
 Eq. (\ref{eq:probeffwavemulti-time}) also explains the normalization in eq. (\ref{eq:defeffwavemulti-time}).\\[0.2cm]
 4. The description of $S_1$ in terms of $\psi_{\rm eff}$ has the same form as the description of $S$ in terms of $\Psi$ (cf. eqs. (\ref{eq:relguid}), (\ref{eq:effwaverelguid}) and eqs. (\ref{eq:crossingprob}), (\ref{eq:probeffwavemulti-time})).\\

\noindent $W_{\rm cond}$ always exists; the effective wave function only exists in certain situations. If the effective wave function exists $W_{\rm cond}$ is pure and given by the effective wave function. The converse is not as obvious and content of the following lemma.

\paragraph{Lemma:} \textit{$W_{\rm cond}(q_1, q_1')$ is pure if and only if for $\Sigma \in \mathcal{F}$, $\Psi(q_1, Q_2(\Sigma))$ can be written as $\Psi^{s_1 s_2}(q_1, Q_2(\Sigma)) = \psi_1^{s_1}(q_1) \psi_2^{s_2}(Q_2(\Sigma))$.}\\

\noindent \textit{Proof:} ``$\Leftarrow$'': Let $\Psi^{s_1 s_2}(q_1, Q_2(\Sigma)) = \psi_1^{s_1}(q_1) \psi_2^{s_2}(Q_2(\Sigma))$. Then, according to eqs. (\ref{eq:wcondrel}) and (\ref{eq:gamman}):
\begin{align}
 {W_{\rm cond}}^{s_1}_{s_1'}(q_1, q_1') ~&=~ \frac{1}{\mathcal{N}} \sum_{s_2} \psi_1^{s_1}(q_1) \psi_2^{s_2}(Q_2(\Sigma)) \left[ \psi_1^\D(q_1') \psi_2^\D(Q_2(\Sigma)) (\gamma n)(Q_2(\Sigma)) \right]_{s_1' s_2}\nonumber\\
        &=~ \frac{1}{\mathcal{N}} \left\{ \sum_{s_2} \psi_2^{s_2}(Q_2(\Sigma)) \left[ \psi_2^\D(Q_2(\Sigma)) (\gamma n)(Q_2(\Sigma)) \right]_{s_2} \right\} \psi_1^{s_1}(q_1) {\psi_1}^\D_{s_1'}(q_1')\nonumber\\
        &\equiv~ \frac{1}{\tilde{\mathcal{N}}} \psi_1^{s_1}(q_1) {\psi_1}^\D_{s_1'}(q_1').
 \label{eq:purewcond}
\end{align}
``$\Rightarrow$'': We split the proof into two steps:\\[0.2cm]
 1. Simplification of the form of $W_{\rm cond}$: Using the manifest Lorentz invariance of the HBD model, we simplify the form of $W_{\rm cond}$ employing the multi-time Lorentz transformation leading to $(\gamma n)(Q_2(\Sigma)) = 1$ (cf. eq. (\ref{eq:multi-timespinorrep}) and below). Note that by virtue of eq. (\ref{eq:multi-timespinorrep}) this transformation leaves a pure $W_{\rm cond}$ pure and a mixed $W_{\rm cond}$ mixed. We obtain: 
 \begin{equation}
   {W_{\rm cond}'}^{s_1}_{s_1'}(q_1, q_1') = \frac{1}{\mathcal{N}} \sum_{s_2} \Psi'^{s_1 s_2}(\tilde{\Lambda}^{-1}(q_1,  Q_2(\Sigma))) \left[ (\Psi')^\D (\tilde{\Lambda}^{-1}(q_1', Q_2(\Sigma)))\right]_{s_1' s_2}.
 \label{eq:wcondreltransformed}
 \end{equation}
 For a flat foliation, this coincides with the definition of the conditional density matrix in the non-relativistic case (cf. eq. (\ref{eq:wcond})).\\[0.2cm]
 2. Indirect proof of ``$\Rightarrow$'' via Schmidt decomposition:
 Assume that for $\Sigma \in \mathcal{F}$ and fixed $q_1$, $\Psi^{s_1 s_2}(q_1, Q_2(\Sigma))$ cannot be written as a tensor product of vectors in $\mathbb{C}^{k_1}$ and $\mathbb{C}^{k_2}$, respectively. Then according to the Schmidt decomposition there exist orthonormal bases $\{ u_1, ..., u_{k_1} \}$ of $\mathbb{C}^{k_1}$ and $\{ v_1, ..., v_{k_2}\}$ of $\mathbb{C}^{k_2}$ such that
 \begin{equation}
  \Psi(q_1,Q_2(\Sigma)) = \sum_{j = 1}^{m} c_j \, u_j \otimes v_j,~~~{\rm in~components:}~~~  \Psi^{s_1 s_2}(q_1,Q_2(\Sigma)) = c_{s_1} \, \delta^{s_1 s_2},
  \label{eq:schmidtdecomp}
 \end{equation}
 where $m = {\rm max}\{ k_1, k_2\}$ and the coefficients $c_j$ are non-negative, uniquely determined by $ \Psi(q_1,Q_2(\Sigma))$ and a number $l \geq 2$ of them is non-zero. Relabel such that these are the coefficients $c_1, ..., c_l$.\\
 Similarly for fixed $q_1'$,
 \begin{equation}
  \Psi(q_1',Q_2(\Sigma)) = \sum_{j = 1}^{m} c_j' \, u_j' \otimes v_j',~~~{\rm in~components:}~~~  \Psi^{s_1' s_2'}(q_1',Q_2(\Sigma)) = c_{s_1'}' \, \delta^{s_1' s_2'},
  \label{eq:schmidtdecomp2}
 \end{equation}
 where $m$ is the same as before, the $u_j'$ and $v_j'$ define orthonormal bases of the respective spaces and we can choose the first $l' \geq 2$ of the $c_j'$ to be non-zero. Then, with the choice of the multi-time Lorentz transformation as before, the conditional density matrix takes the following form:
 \begin{equation}
  {W_{\rm cond}}^{s_1}_{s_1'}(q_1,q_1') = \frac{1}{\mathcal{N}} \sum_{s_2 = s_2'} c_{s_1} c_{s_1'}' \delta^{s_1 s_2} \delta_{s_1' s_2'} = \frac{1}{\mathcal{N}} c_{s_1} c_{s_1'}' \delta^{s_1}_{s_1'}.
  \label{eq:wschmidt}
 \end{equation}
 Thus, as a diagonal matrix with ${\rm min}\{l, l'\} \geq 2$ non-zero entries, ${W_{\rm cond}}^{s_1}_{s_1'}(q_1,q_1')$ is not pure.
\begin{flushright}
  $\Box$\\[0.2cm]
\end{flushright}

\noindent The lemma shows that the mathematical structure in the definition of the effective wave function (cf. eq. (\ref{eq:effwavemulti-time})) follows from $W_{\rm cond}$ being pure. If in addition the ``macroscopic disjointness'', the key aspect to earn $\psi_{\rm eff}$ the attribute ``effective'', is given, the effective wave function is indeed the wave function that is uniquely determined by $W_{\rm cond}$.

\section{Indistinguishable Particles}
Particles with spin are usually thought of as being indistinguishable. Contrary to what one may think, this poses no problem for BM (cf. \cite{topology}). One only has to recognize that the appropriate configuration space in this case is the set of unordered configurations, i.e. the subsets of $\mathbb{R}^3$ with $N$ elements: $\,^N \mathbb{R}^3 \equiv \{ S \subset \mathbb{R}^3: |S| = N\}$. This space is topologically nontrivial and the analysis of BM on this space leads to the familiar distinction between bosons and fermions (see e.g. \cite{topology}). The description of the spin bundle on this configuration space becomes, however, a bit technical. Since such technicalities do not yield more insight into the autonomous subsystem description, we adopt a pragmatic point of view here: One may use an arbitrary labeling of the particles (and thus the ordered tensor product of spinors) and apply the corresponding (anti)symmetrization postulate. The crucial point is that the constructions in the definitions of $W_{
\rm cond}$ and $\psi_{\rm eff}$ in fact commute with permutations of the particle labels. Thus, one may apply them without changes. In particular, the distinction between environment particles and subsystem particles is not based on particle labels, but on the fact that some set of particles, e.g. particles belonging to a certain region $R \subset \Sigma$ of a hypersurface $\Sigma \in \mathcal{F}$, has actual configuration $\{ X(\Sigma), Y(\Sigma),..., Z(\Sigma)\} \subset R$.

\section{Outlook}
In this paper we have derived a subsystem description for the HBD model. By construction, the subsystem's variables (the conditional density matrix and the effective wave function) have statistical meaning on the spacetime structure given by the foliation $\mathcal{F}$.\\ 
This framework should be taken as the starting point for a further study of an effective relativistic ``measurement formalism'' which in turn should lead to a rigorous justification of the usual quantum formalism (as far as it exists for relativistic quantum theories). A thorough discussion of the measurement formalism as arising from non-relativistic BM has been achieved in \cite{operators,qpwqp}.\\
To appreciate the quest, note that a physical experiment with its state of motion defines an equal time hypersurface $\mathcal{S}$ in general not belonging to the foliation. That is, such a hypersurface $\mathcal{S}$ is not part of the space time structure defined by $\mathcal{F}$. Of course, the main interest lies in the statistics of outcomes of ``measurements'' of the subsystems on such a ``query hypersurface''. In order to obtain these statistics, one has to relate the ``$|\psi|^2$-probability'' formula holding only on leaves of $\mathcal{F}$ to the usual formalism of operator-observables on query hypersurfaces. An analysis related in spirit was performed in \cite{berndl} for a particular limiting case of Lorentz invariance and spin-less particles.

\subsection*{Acknowledgments} The authors would like to thank Nicola Vona for helpful discussions. This work was partially supported by COST Action MP1006. M.L. gratefully acknowledges financial support by the German National Academic Foundation.

\end{document}